\documentclass[sigconf]{acmart}

\usepackage{xcolor}
\usepackage{times}
\usepackage{soul}
\usepackage{url}
\usepackage{hyperref}
\usepackage{caption}
\usepackage{amsmath}
\usepackage{amsthm}
\usepackage{tabularray}

\usepackage{booktabs}
\usepackage{algorithm}
\usepackage{algpseudocode}

\usepackage[switch]{lineno}
\usepackage[utf8]{inputenc}
\usepackage{longtable}
\usepackage{enumitem}
\usepackage{tcolorbox}
\usepackage{multirow}

\usepackage[utf8]{inputenc}
\usepackage{CJKutf8} 
\usepackage{tcolorbox}
\usepackage{xcolor}
\definecolor{newsColor}{rgb}{0.0, 0.5, 0.0}
\definecolor{financeColor}{rgb}{0.0, 0.0, 1.0}
\definecolor{stockPriceColor}{rgb}{1.0, 0.0, 0.0}
\tcbuselibrary{skins}
\usepackage{mathtools}

\AtBeginDocument{%
  }

\setcopyright{acmlicensed}
\copyrightyear{2024}
\acmYear{2024}
\acmDOI{XXXXXXX.XXXXXXX}

\acmConference[ICAIF '24]{ACM International Conference on AI in Finance}{November 14-16,2024}{New York, NY, USA}
\acmISBN{978-1-4503-XXXX-X/18/06}

\begin{document}

\title{FinRobot: AI Agent for Equity Research and Valuation with  \\ Large Language Models}

\author{Tianyu Zhou}
\authornote{These authors contributed equally to this research.}
\affiliation{%
  \institution{AI4Finance Foundation}
  \city{New York}
  \country{USA}
}
\email{tianyu_zhou1@brown.edu}

\author{Pinqiao Wang}
\authornotemark[1]
\affiliation{%
  \institution{AI4Finance Foundation}
  \city{New York}
  \country{USA}
  }
\email{pw2594@columbia.edu}

\author{Yilin Wu}
\authornotemark[1]
\affiliation{%
  \institution{AI4Finance Foundation}
  \city{New York}
  \country{USA}}
\email{yilin002@e.ntu.edu.sg}

\author{Hongyang Yang}
\authornote{Corresponding Author}
\affiliation{%
 \institution{AI4Finance Foundation}
 \city{New York}
  \country{USA}}
 \email{hy2500@columbia.edu}




\begin{abstract}

As financial markets grow increasingly complex, there is a rising need for automated tools that can effectively assist human analysts in equity research, particularly within sell-side research. While Generative AI (GenAI) has attracted significant attention in this field, existing AI solutions often fall short due to their narrow focus on technical factors and limited capacity for discretionary judgment. These limitations hinder their ability to adapt to new data in real-time and accurately assess risks, which diminishes their practical value for investors.

This paper presents FinRobot, the first AI agent framework specifically designed for equity research. FinRobot employs a multi-agent Chain of Thought (CoT) system, integrating both quantitative and qualitative analyses to emulate the comprehensive reasoning of a human analyst. The system is structured around three specialized agents: the Data-CoT Agent, which aggregates diverse data sources for robust financial integration; the Concept-CoT Agent, which mimics an analyst’s reasoning to generate actionable insights; and the Thesis-CoT Agent, which synthesizes these insights into a coherent investment thesis and report. FinRobot provides thorough company analysis supported by precise numerical data, industry-appropriate valuation metrics, and realistic risk assessments. Its dynamically updatable data pipeline ensures that research remains timely and relevant, adapting seamlessly to new financial information. Unlike existing automated research tools, such as CapitalCube and Wright Reports, FinRobot delivers insights comparable to those produced by major brokerage firms and fundamental research vendors. We open-source FinRobot at \url{https://github. com/AI4Finance-Foundation/FinRobot}.


\end{abstract}

\ccsdesc[300]{Computing methodologies~Natural language processing}

\keywords{AI-agent, Large Language Models, Equity Research, Financial Analysis, Chain of Thought}


\maketitle

\section{Introduction}
Financial analysis is a cornerstone of decision-making in the financial services industry, essential for guiding investment choices and helping institutions and individuals navigate complex markets \cite{abarbanell1997fundamental,greenwald2004value,penman2010accounting,berman2013financial,subramanyam2014financial}. Within the scope of financial analysis, equity research holds particular importance, especially in sell-side research departments of major investment banks and brokerages. These departments rely on highly specialized analysts to produce detailed stock reports that provide valuable insights for investment decision-making, influencing strategies for both institutional and retail investors. However, producing high-quality equity research is a labor-intensive process that requires not only an in-depth understanding of quantitative models but also industry-specific expertise to contextualize raw data effectively.

With advances in artificial intelligence (AI) and Large Language Models (LLMs) \cite{medhat2014sentiment,zhang2023fingptrag,henrique2019literature,nabipour2020deep,kumar2022systematic,jiang2021applications,brown2020language,wu2023bloomberggpt,yang2023fingpt,kim2024financial}, there is growing interest in automating elements of the equity research process. While existing AI tools—such as CapitalCube, Wright Reports, and MarketGrader—offer automated insights, they often fall short in replicating the comprehensive, nuanced analysis characteristic of human analysts. These tools tend to focus heavily on technical factors and rely on simplified models, which limits their ability to incorporate discretionary judgment and qualitative assessments, both of which are crucial for robust equity research.


In this paper, we introduce FinRobot, an AI-powered equity research platform that bridges key gaps in automated analysis. Using a multi-agent financial Chain of Thought (CoT) system, FinRobot combines discretionary judgment with quantitative analysis, meeting the standards of leading institutions like JPMorgan and UBS. Its real-time data pipeline ensures timely, relevant insights, while its realistic risk assessments set it apart from existing tools. By blending academic rigor with practical applications, FinRobot marks a significant advancement in AI-driven sell-side research.

The primary contributions of this paper are as follows:
\begin{itemize}
    \item \textbf{First AI Agent for Equity Research Using a Multi-Agent Chain of Thought (CoT) System}: FinRobot introduces the first AI-driven approach for equity research with a multi-layer CoT framework. Through specialized agents (Data-CoT, Concept-CoT, Thesis-CoT), it simulates the analytical depth and narrative skills of human analysts across data processing, concept generation, and report synthesis.

    \item \textbf{Blending Discretionary Judgment with Real-Time Data and New Evaluation Metrics}: FinRobot integrates quantitative and qualitative analysis, mimicking human judgment. It features a real-time data pipeline for up-to-date insights and employs \textit{Accuracy}, \textit{Logicality}, and \textit{Storytelling} metrics to evaluate report quality.

    \item \textbf{Open-Source Platform to Democratize Financial AI}: By being open-source, FinRobot makes advanced AI tools accessible to the finance sector, fostering collaboration and innovation across the AI and financial communities.
\end{itemize}

\begin{figure*}[h]
    \centering
    \includegraphics[width=0.77\textwidth]{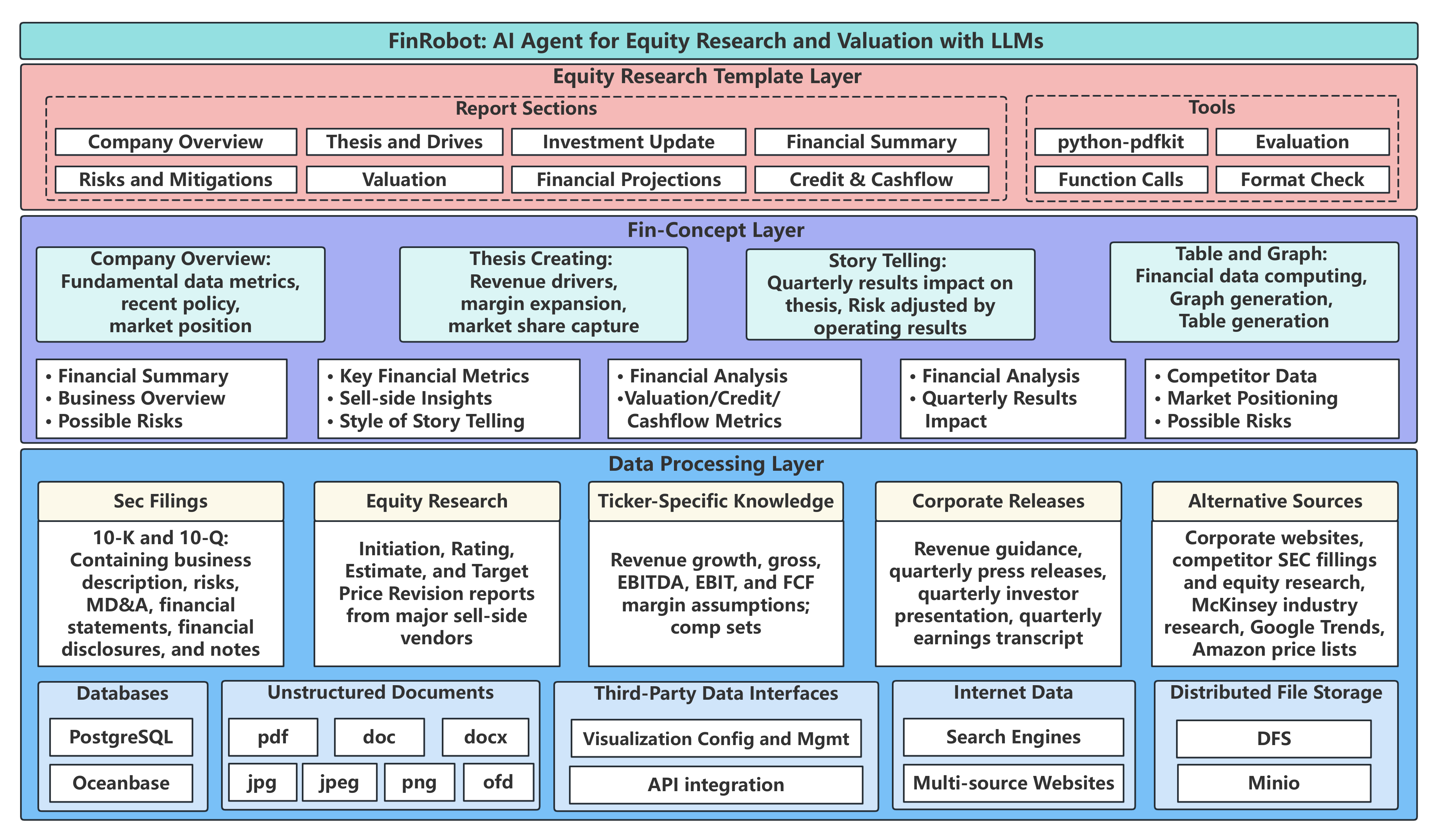}
\caption{Overall Framework of FinRobot.}
\label{fig:framework}
\vspace{-2mm}
\end{figure*}

\section{Related Works}

\subsection{LLMs and Financial Applications}
Large Language Models (LLMs) have gained significant attention within AI research due to their capabilities in natural language processing and data interpretation \cite{zhao2023survey,radford2018improving,brown2020language}. In the context of financial analysis, several studies have investigated the application of LLMs for automating tasks that are traditionally performed by human analysts \cite{nie2024survey}. For example, tasks such as sentiment analysis \cite{medhat2014sentiment,sohangir2018big,huang2023finbert,zhang2023fingptrag} and market prediction \cite{henrique2019literature,nabipour2020deep,kumar2022systematic,jiang2021applications} have been primary areas of focus for financial AI applications. However, while LLMs have shown promise in automating parts of the financial analysis process, they often lack the real-time adaptability and sector-specific expertise necessary for precise equity research. As a result, their application can be limited in environments that demand rapid updates and a nuanced understanding of financial metrics.

\subsection{AI Agents and CoT in Financial Analysis}
AI agents leveraging multi-agent collaboration frameworks have emerged as powerful tools in financial analysis, particularly for replicating human decision-making processes. Systems like FinAgent \cite{zhang2024finagent} and FinMem \cite{yu2023finmem} exemplify how Large Language Models (LLMs) can enhance trading strategies by utilizing real-time market data for more informed decision-making. Notably, LLMs have demonstrated the capability to match or even surpass human analysts in accurately predicting earnings trends. A key innovation in this space is the use of Chain-of-Thought (CoT) prompting, which enables models to emulate human reasoning steps, thereby improving predictive precision. By employing CoT, models such as GPT-4 can dissect financial ratios and trends, generating insightful analysis that aids financial decision-making \cite{wei2022chain,kim2024financial}.

Building on this foundation, FinRobot incorporates a CoT framework tailored specifically for equity research, replicating the nuanced reasoning of a financial analyst. This design not only provides depth in analysis but also enhances agility, allowing FinRobot to adapt to diverse financial contexts and rapidly changing data. By advancing the application of AI agents with integrated CoT systems, FinRobot addresses the limitations of traditional AI-based tools, boosting both the accuracy and practical relevance of equity research in the financial sector.

\section{Methodology}
\subsection{Overview} 
As delineated in Fig. \ref{fig:framework}, the overall framework of FinRobot is designed around a multi-layer Chain of Thought (CoT) framework, which structures the financial equity research process to emulate the logic and narrative skills of a seasoned sell-side analyst. This architecture not only enhances the coherence and depth of analysis but also allows FinRobot to produce reports that are both insightful and immediately usable.

The multi-layer CoT framework decomposes the complex task of financial analysis into sequential, manageable steps, each dedicated to a specific aspect of equity research. By doing so, FinRobot achieves a deeper understanding of financial data and simulates human-like reasoning across multiple dimensions. This framework is built upon three primary layers, each of which is aligned with a specific CoT agent:

\begin{enumerate}
\item \textbf{Data Processing Layer (Data-CoT Agent):} 
\begin{itemize}
\item The Data Processing Layer is responsible for gathering and preparing financial data from various sources, such as SEC filings, earnings call transcripts, and alternative data streams. This layer ensures that the data is accurate, comprehensive, and ready for analysis, forming a solid foundation for downstream processes.
\item Within this layer, the Data-CoT agent handles the integration and extraction of essential financial metrics, converting raw data into structured summaries. The agent focuses on capturing both quantitative metrics and qualitative insights, setting up a baseline that supports the analysis performed in the subsequent layers.
\end{itemize}

\item \textbf{Financial Concept Layer (Concept-CoT Agent):} 
\begin{itemize}
\item The Financial Concept Layer interprets the processed data to develop actionable financial insights. This layer is dedicated to understanding key metrics, such as revenue projections, EBITDA trends, and market positioning, and provides the conceptual basis for a company's financial performance evaluation.
\item The Concept-CoT agent within this layer mimics the thought process of a human analyst by contextualizing data. It performs tasks such as competitive analysis and sentiment evaluation, applying financial models and simulating analyst-driven inquiry to derive a well-rounded view of the company’s prospects and challenges.
\end{itemize}

\begin{figure*}[h]
    \centering
    \includegraphics[width=0.8\textwidth]{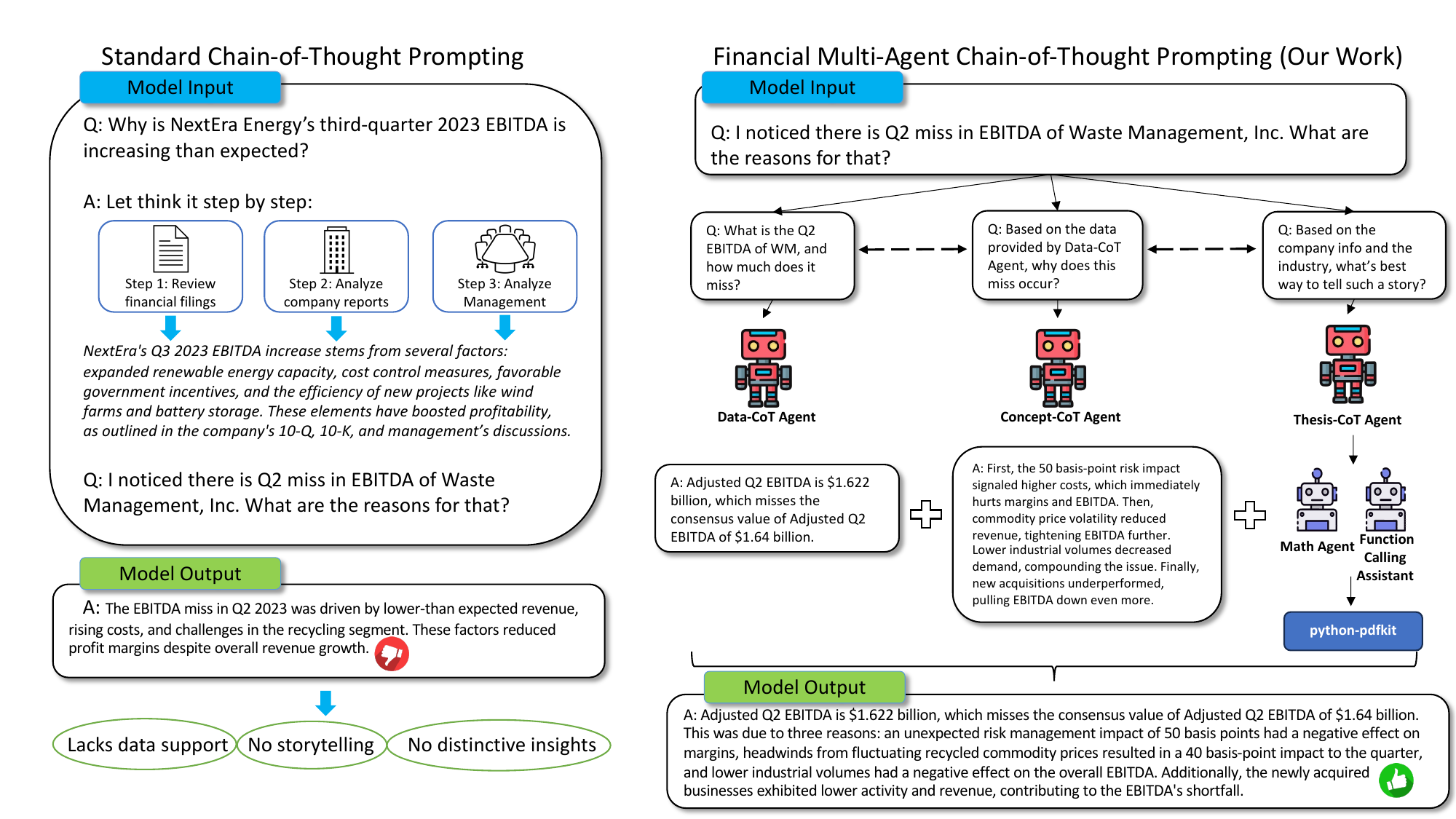}
    \caption{Standard Chain-of-Thought (CoT) vs Financial Multi-Agent Chain-of-Thought (CoT)}
    \label{fig:finrobot_cot}
    \vspace{-2mm}
\end{figure*}

\item \textbf{Equity Research Template Layer (Thesis-CoT Agent):} 

\begin{itemize}
\item The Equity Research Template Layer compiles and synthesizes the insights generated from the preceding layers into a structured equity research report. This layer ensures that the final output includes a coherent investment thesis, complete with financial projections, risk assessments, and valuation modeling.
\item Operating within this layer, the Thesis-CoT agent organizes the findings into a report format commonly used in sell-side research. It combines quantitative data with qualitative judgment to provide a final recommendation (e.g., buy, hold, sell). By integrating elements of narrative-building and strategic assessment, the Thesis-CoT agent ensures the report is both compelling and aligned with professional equity research standards.
\end{itemize}

\end{enumerate}

\subsection{Data Processing Layer}

The \textit{Data Processing Layer} forms the foundation of FinRobot's methodology, focusing on the collection and initial processing of financial data. The primary objective of this layer is to ensure that all data inputs are accurate, comprehensive, and well-structured, as they provide the basis for subsequent analyses. This layer is designed to manage diverse data sources, integrating both structured and unstructured data to build a reliable dataset for in-depth financial evaluation.

\textbf{Data-CoT Agent:} The Data-CoT agent plays a crucial role in the Data Processing Layer, tasked with collecting raw financial data from multiple external sources, including SEC filings, corporate releases, and earnings call transcripts. By extracting and processing this data, the agent ensures that downstream analysis is built on accurate and relevant information. The Data-CoT agent generates mid-level summaries, including both numerical metrics and textual insights, laying the groundwork for advanced analysis in later stages. Key financial metrics calculated by the Data-CoT agent are outlined in Table ~\ref{tab:Financial_Formulas}. These metrics, such as revenue growth, contribution profit, and EBITDA margin, offer a detailed view of the company's financial performance. These calculations form the basis of FinRobot’s ability to track financial trends over time and evaluate company performance.

\textbf{Data Source:} FinRobot's data processing framework is built upon a diverse range of data sources that ensure comprehensive and accurate financial analysis.
\begin{itemize}
    \item \textbf{Database:} FinRobot utilizes structured databases such as Oceanbase and PostgreSQL for efficient storage, retrieval, and management of structured financial data, supporting complex querying and transactional operations.
    \item \textbf{Unstructured Documents:} To handle a wide variety of document formats, FinRobot processes unstructured documents including PDF, DOC, DOCX, JPEG, and PNG files, extracting textual and visual information necessary for comprehensive financial analysis.
    \item \textbf{Third-Party Data Interfaces:} The system integrates third-party data interfaces for visualization configuration and management, along with API integration, enabling seamless access to external data sources and real-time updates.
    \item \textbf{Internet Data:} FinRobot leverages internet data, incorporating insights from search engines and various multi-source websites to gather up-to-date information on market trends, industry developments, and competitive landscape.
    \item \textbf{Distributed File Storage:} Distributed file storage solutions like DFS and Minio provide scalable and resilient storage for large volumes of data, ensuring high availability and robustness in data management across the system.
\end{itemize}



\begin{table*}
\begin{center}
\begin{tabular}{|p{7.5cm}|p{9.5cm}|}
\hline
\textbf{Formula} & \textbf{Description} \\
\hline
$\text{Revenue Growth} = \frac{\text{Revenue}_{\text{current}} - \text{Revenue}_{\text{previous}}}{\text{Revenue}_{\text{previous}}}$ 
& Calculate the growth in revenue compared to the previous period. \\
\hline
$\text{Revenue Growth Projection} = \text{Revenue Growth}_{\text{previous}} + 1\%$ 
& Projected revenue growth with a 1\% increase over the previous growth. \\
\hline
$\text{Contribution Profit} = \text{Revenue} - \text{Operating Expense}$ 
& Contribution profit calculated by subtracting operating expenses from revenue. \\
\hline
$\text{Contribution Margin} = \frac{\text{Contribution Profit}}{\text{Revenue}}$ 
& Measure the contribution margin as a percentage of revenue. \\
\hline
\small $\text{Contribution Margin Projection} = \text{Contribution Margin}_{\text{previous}} + 0.5\%$ 
& Projected contribution margin with a 0.5\% increase over the previous margin. \\
\hline
$\text{SG\&A Margin} = \frac{\text{SG\&A}}{\text{Revenue}}$ 
& SG\&A expenses as a percentage of revenue. \\
\hline
$\text{EBITDA} = \text{Contribution Profit} - \text{SG\&A}$ 
& Calculate EBITDA by subtracting SG\&A from the contribution profit. \\
\hline
$\text{EBITDA Margin} = \frac{\text{EBITDA}}{\text{Revenue}}$ 
& EBITDA as a percentage of revenue. \\
\hline
$\text{CAGR} = \left( \frac{\text{EV}}{\text{BV}} \right)^{\frac{1}{n}} - 1 \times 100$ 
& Calculate the Compound Annual Growth Rate, where $\text{EV}$ is the ending value, $\text{BV}$ is the beginning value, and $n$ is the number of years. \\
\hline
$\text{Enterprise Multiple} = \frac{\text{EV}}{\text{EBITDA}}$ 
& The Enterprise Multiple, where $\text{EV}$ is the Enterprise Value, and $\text{EBITDA}$ is earnings before interest, taxes, depreciation, and amortization. \\
\hline
\end{tabular}
\end{center}
\caption{Financial Formulas}
\label{tab:Financial_Formulas}
\vspace{-4mm}
\end{table*}

\textbf{SEC Filings}: A primary source of financial data comes from SEC filings such as 10-K and 10-Q reports. These documents provide detailed insights into a company's financial performance, risks, business descriptions, management strategies, and regulatory disclosures. FinRobot automatically extracts relevant data (e.g. revenue, operating cost, SG\&A) from these filings, focusing on key areas like income statements, balance sheets, cash flow statements, and the Management Discussion \& Analysis (MD\&A) section. Using the original collected data, FinRobot calculates some important metrics (e.g. revenue growth, contribution profit, contribution margin, SG\&A margin, EBITDA, EBITDA margin) used for equity research. 

\textbf{Equity Research Reports}: FinRobot leverages sell-side analysts' reports that offer financial estimates, target price revisions, and stock ratings. These reports serve as benchmarks for model validation, enabling FinRobot to align its projections with market expectations. By identifying discrepancies across various analyst reports, FinRobot refines its risk assessments and creates a balanced perspective on stock performance.

\textbf{Corporate Releases}: Quarterly earnings releases and investor presentations are integrated into the data pipeline. These documents provide real-time updates on financial performance and forward guidance, which are crucial for updating the financial models. FinRobot’s dynamic data integration ensures that new developments, such as changes in revenue guidance or profitability metrics, are immediately reflected in the analysis.

\textbf{Earning Call Transcript}: FinRobot incorporates earnings call transcripts into its data pipeline to capture recent insights from company executives. These transcripts provide context on quarterly results, strategic plans, and future outlooks, which go beyond mere numbers. By swiftly integrating updates from earnings calls, FinRobot maintains an accurate view of a company’s financial health and strategic direction, adapting to shifts in market trends and management commentary.

\textbf{Alternative Data Sources}: Beyond traditional financial metrics, FinRobot taps into alternative data sources like competitor filings, industry research, Google Trends, and consumer sentiment from platforms such as Amazon. These additional data points offer contextual layers that enhance FinRobot's understanding of market dynamics and competitive positioning.

\subsection{Fin-Concept Layer}

The \textit{Fin-Concept Layer} is where raw data is processed into meaningful financial concepts and projections. This layer transforms the collected data into actionable insights, focusing on financial metrics and future projections. The layer includes the \textit{Concept-CoT} agent, allowing FinRobot to simulate an analyst's thought process and respond to complex financial queries as shown in Fig. \ref{fig:finrobot_cot}.

\textbf{Concept-CoT Agent}: The \textit{Concept Chain-of-Thought (CoT)} layer simulates the reasoning process of a human analyst. Once raw data is transformed into financial metrics by data CoT, FinRobot uses the Concept-CoT to ask and answer critical questions that drive deeper analysis. For example, FinRobot may question the key drivers of revenue growth, how a company’s margins compare to its competitors, and what potential risks could affect its future performance. The Concept-CoT framework ensures that each financial metric is analyzed in isolation and contextualized within the broader market and industry environment.

\textbf{Revenue Projections}: A core function of the Fin-Concept Layer is generating revenue projections. This is done using a combination of historical data and forward-looking estimates. For companies in the industrial sector, for example, revenue is typically modeled as the product of volume and price. FinRobot accounts for backlog data as an indicator of future sales, while factoring in pricing trends, inflationary pressures, and competitive dynamics to refine its revenue forecasts. By adjusting for these variables, FinRobot can provide both optimistic and conservative revenue scenarios.

\textbf{EBITDA and Margin Trends}: FinRobot also conducts detailed EBITDA (Earnings Before Interest, Taxes, Depreciation, and Amortization) calculations, a key measure of a company’s operational profitability. These calculations are adjusted for one-time items, giving a more accurate picture of ongoing profitability. In addition, FinRobot tracks both gross and EBITDA margins, analyzing trends over time and comparing them to industry benchmarks. Margin trends are critical for assessing the efficiency and sustainability of a company’s operations, providing investors with a clearer picture of long-term profitability.

\textbf{ROIC and WACC Calculations}: The Fin-Concept Layer also calculates Return on Invested Capital (ROIC) and Weighted Average Cost of Capital (WACC). ROIC measures the efficiency of a company’s capital deployment, while WACC calculates the cost of capital (both debt and equity). These two metrics are integral to FinRobot’s valuation models, as they help determine whether a company is generating sufficient returns to justify its capital expenditures. FinRobot uses these calculations to build Discounted Cash Flow (DCF) models, which provide a target price for the stock.

\textbf{Financial Query}: FinRobot also responds to specific financial queries that investors or analysts might pose. These queries may relate to a company’s performance relative to its peers, the impact of market changes on profitability, or the likelihood of margin expansion over the next fiscal year. FinRobot provides well-reasoned answers grounded in quantitative and qualitative data by processing these queries through the Concept-CoT framework.

\subsection{Equity Research Template Layer}

The final step in FinRobot’s methodology is the \textit{Equity Research Template Layer}, where the outputs from the Fin-Concept Layer are structured into a standardized research report. This layer ensures that the research report follows the same format and rigor as those produced by traditional sell-side research departments. Within this layer, the \textit{Thesis-CoT} process helps FinRobot develop a clear investment thesis for each stock it analyzes.

\textbf{Thesis-CoT Agent}: The \textit{Thesis Chain-of-Thought (CoT)} agent is responsible for constructing the final investment recommendation. This process involves synthesizing the outputs from the Concept-CoT layer and using them to build a coherent argument for why a stock should be rated as a buy, hold, or sell. The Thesis-CoT considers both quantitative data and qualitative factors, such as management quality, market sentiment, and competitive threats. This ensures that FinRobot’s investment thesis is holistic, considering the financial and strategic aspects of a company’s performance.

\textbf{Thesis Development}: The central output of the Equity Research Template Layer is a well-defined investment thesis. This thesis explains the stock’s key revenue drivers, competitive positioning, and growth prospects. The thesis is supported by data from the Fin-Concept Layer, ensuring that all conclusions are backed by quantitative analysis. The thesis may also include a qualitative discussion of market risks, regulatory challenges, or strategic initiatives that could influence the company’s future performance.

\textbf{Risk}: The Equity Research Template Layer also incorporates a comprehensive risk analysis, identifying potential challenges that may impact the company's future performance and investment appeal. This risk analysis helps investors understand the factors that could affect the company’s valuation and growth trajectory, enabling more informed investment decisions based on both the opportunities and challenges the company may face.

\textbf{Valuation and Financial Projections}: In addition to the thesis, FinRobot provides detailed financial projections, including revenue, EBITDA, and margin forecasts for the next several fiscal years. These projections are integrated into valuation models, such as the Discounted Cash Flow (DCF) or Price-to-Earnings (P/E) ratio models. The report also includes target prices for the stock based on these projections, offering investors a clear recommendation (e.g., buy, hold, or sell) based on the underlying financial analysis.

\textbf{Competitor Analysis}:
FinRobot further enhances its investment thesis by providing a detailed competitor analysis. This section evaluates the company's performance in the context of its industry peers, using key metrics such as revenue growth, gross margin, EBITDA margin, and SG\&A expense margin. By benchmarking these metrics against competing companies, FinRobot offers insights into the company’s operational efficiency, profitability, and market positioning. This analysis equips investors with a clearer understanding of the company’s competitive advantages and potential vulnerabilities, enabling more informed investment decisions.

\section{Experiments}  

\subsection{Task Description}
FinRobot is capable of generating equity research reports across various sectors. For this demonstration, we focused on the energy sector, using Waste Management, Inc. as an example. As a leading provider of waste management and environmental services in North America, Waste Management offers collection, recycling, and disposal solutions for a wide range of customers, from residential to industrial, with a strong emphasis on sustainability.

The FinRobot-generated report offers a comprehensive overview, covering Waste Management’s financial performance, investment potential, valuation, and risk factors. It includes tables and charts that highlight key financial metrics, stock performance, and industry comparisons. This detailed analysis delivers accurate, data-driven insights and strategic guidance for investors, reflecting current market trends and external influences.

The full equity research report for Waste Management, Inc. generated by FinRobot can be found in the appendix.

\begin{table}[h!]
\centering
\begin{tabular}{|c|c|c|c|}
\hline
\textbf{Reviewer} & \textbf{Accuracy} & \textbf{Logicality} & \textbf{Storytelling} \\
\hline
Reviewer 1 & 10 & 10 & 10 \\
\hline
Reviewer 2 & 10 & 9 & 8 \\
\hline
Reviewer 3 & 10 & 9 & 8 \\
\hline
Reviewer 4 & 9 & 9 & 7 \\
\hline
Reviewer 5 & 9 & 10 & 7 \\
\hline
Reviewer 6 & 9 & 9 & 10 \\
\hline
Reviewer 7 & 10 & 9.5 & 8.5 \\
\hline
\end{tabular}
\caption{Scores for Equity Research Report by Seven Reviewers}
\label{table:scores}
\vspace{-3mm}
\end{table}

\begin{figure}[ht!]
\centering
\begin{tcolorbox}[
  enhanced,
  colback=white,
  boxrule=0.5pt,
  arc=4pt,
  left=6pt,
  right=6pt,
  top=6pt,
  bottom=6pt
]
\small 
\textbf{Instruction:} You are an experienced financial analyst. Your task is to evaluate the equity research report on Waste Management based on three specific dimensions: accuracy, logicality, and storytelling ability.
Your answer format should be as follows: \\

{[Accuracy] Score:}\\
1. ...

{[Logicality] Score:}\\
1. ...

{[Storytelling] Score:}\\
1. ...\\

\textbf{Information:}\\
a. Accuracy : Assess how accurate the information in the report is, focusing on the precision of financial data, adherence to known facts, and the overall credibility of the presented analysis. Provide a score from 0 to 10, with 10 being extremely accurate.\\
b. Logicality : Evaluate the report’s logical structure, looking at the clarity of the flow from one section to the next, the coherence of arguments, and the logical progression of financial analysis. Provide a score from 0 to 10, with 10 representing a perfectly logical and well-organized report.\\
c. Storytelling Ability : Judge the storytelling aspect, considering whether the report is engaging, provides meaningful context, and effectively communicates the importance of financial data in an insightful manner. Provide a score from 0 to 10, with 10 being highly engaging and insightful.\\

\end{tcolorbox}
\vspace{-2mm}
\caption{Evaluation Prompt Template}
\label{fig:prompt_template}
\vspace{-3mm}
\end{figure}

\begin{figure}[ht!]
\centering
\begin{tcolorbox}[
  enhanced,
  colback=white,
  boxrule=0.5pt,
  arc=4pt,
  left=6pt,
  right=6pt,
  top=6pt,
  bottom=6pt
]
\small 
\textbf{Comments generated by GPT-4}

\textbf{[Accuracy] 9:}\\
The report presents a high level of detail regarding Waste Management's financials, including key metrics such as revenue, profit margins, and projections. The data appears well-researched and precise, though some minor updates on recent financials could further enhance its accuracy. Overall, the financial figures and analyses align with known market data, contributing to the report's reliability.\\

\textbf{[Logicality] 8:}\\
The report follows a logical structure, beginning with an overview and then diving into various financial metrics. Each section builds upon the previous one, allowing the reader to follow the financial narrative smoothly. However, a few transitions between sections could be clearer, especially when moving from qualitative to quantitative analyses. Some sections might benefit from additional context or explanation, particularly where assumptions are made.\\

\textbf{[Storytelling] 7:}\\
While the report is data-driven, it could benefit from a more engaging narrative. Currently, it reads as a well-organized list of facts rather than a compelling story. Adding context to explain why certain metrics are important to Waste Management's performance would enhance the storytelling aspect.\\
\end{tcolorbox}
\vspace{-3mm}
\caption{Comments from GPT-4}
\label{fig:gpt_ans1}
\vspace{-3mm}
\end{figure}

\begin{table*}[h]
\centering
\begin{tabular}{|c|p{5cm}|p{5cm}|p{5cm}|}
\hline
\textbf{Score} & \textbf{Accuracy} & \textbf{Logicality} & \textbf{Storytelling} \\ \hline
10 & Perfect accuracy, no errors or inconsistencies. & Exceptionally logical flow, perfectly structured. & Exceptionally engaging, flawless storytelling. \\ \hline
9 & Highly accurate with only minor errors. & Very logical, well-structured with minor flaws. & Strong storytelling, minor areas for improvement. \\ \hline
8 & Mostly accurate, a few minor inconsistencies. & Logical and mostly well-structured, minor issues. & Engaging and well-paced, slight room for improvement. \\ \hline
7 & Accurate, with a few minor issues. & Generally logical with clear structure. & Good storytelling, clear and engaging. \\ \hline
6 & Satisfactory accuracy, some minor errors present. & Mostly logical but may lack depth in reasoning. & Clear storytelling, but lacks depth or consistency. \\ \hline
5 & Fairly accurate but noticeable issues. & Basic logical structure, some gaps in flow. & Basic storytelling, lacks engagement. \\ \hline
4 & Somewhat accurate but may contain errors. & Lacks consistent logical structure, some confusion. & Unengaging storytelling, basic and inconsistent. \\ \hline
3 & Contains frequent inaccuracies. & Poor logical flow, structure is disjointed or unclear. & Storytelling is unclear, lacks engagement. \\ \hline
2 & Inadequate accuracy, many errors. & Very little logical structure, mostly confusing. & Poor storytelling, lacks clarity and coherence. \\ \hline
1 & Highly inaccurate, numerous errors. & Little to no logical flow, very confusing structure. & Extremely poor storytelling, difficult to follow. \\ \hline
0 & Completely inaccurate, no correct information. & No logical structure, completely incoherent. & No storytelling structure, entirely confusing. \\ \hline
\end{tabular}
\caption{Evaluation Criteria for Accuracy, Logicality, and Storytelling in Equity Research Reports}
\label{evaluation_criteria}
\end{table*}

\begin{figure*}[h]
    \centering
    \includegraphics[width=0.8\textwidth]{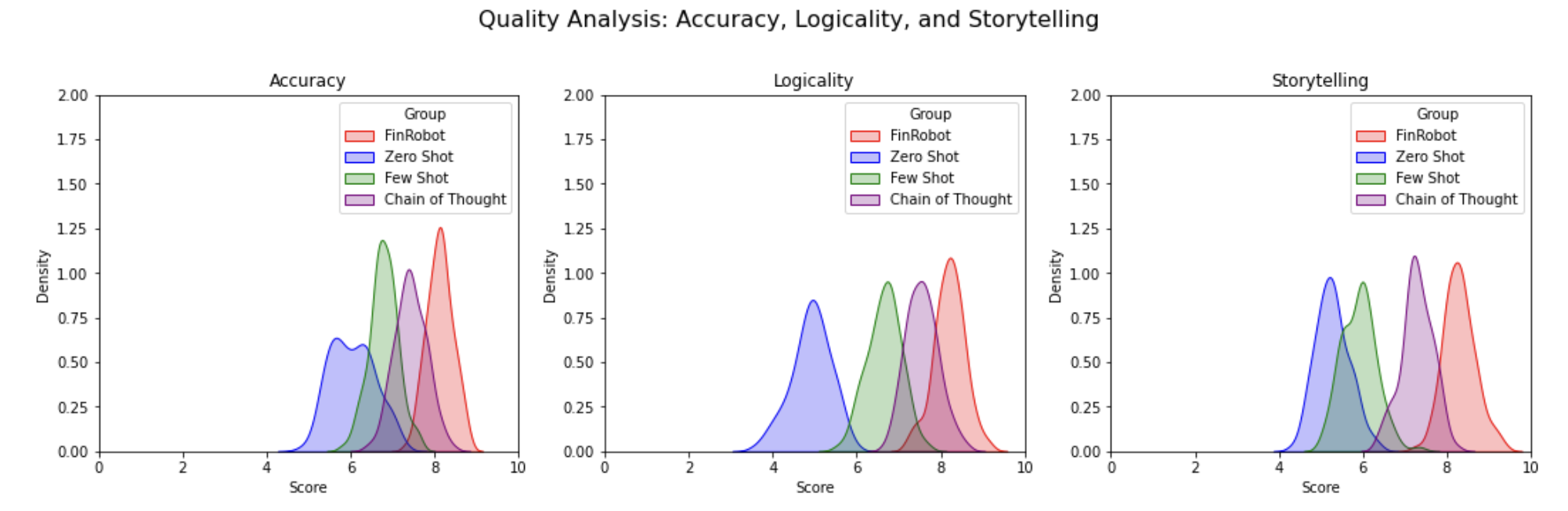}
    \caption{Quality Analysis: Accuracy, Logicality, and Storytelling}
    \label{fig:quality_analysis}
    \vspace{-2mm}
\end{figure*}   

\subsection{Implementation Details}
In line with the methodology outlined, FinRobot was guided to generate an equity research report on Waste Management, Inc. The process begins with the Data-CoT layer, where FinRobot collects raw data from sources such as SEC filings and corporate releases, compiling all relevant information about the company. This layer processes and organizes the data into mid-level summaries, which serve as the foundation for further analysis.

Next, the Concept-CoT layer emulates the reasoning process of a human analyst by evaluating the compiled data to address key questions regarding Waste Management’s financial performance and earnings projections. Through this layer, FinRobot generates insights on operational metrics, market positioning, and potential future performance.

Finally, the Thesis-CoT layer formats these insights into a cohesive and visually appealing report. This layer ensures that the report adheres to industry standards, with a professional layout that is both informative and engaging for investor presentations. This structured approach enables FinRobot to deliver accurate, data-driven insights, maintaining a high standard of clarity and presentation quality.

\subsection{Evaluation}

\subsubsection{Expert Review} To evaluate FinRobot's performance, we engaged a panel of investment banking analysts to review and rate the generated report. Experts were asked to score the report on a scale of 0 to 10 across three key dimensions: accuracy, logical coherence, and storytelling ability, with 10 representing the highest possible score. The detailed evaluation criteria are outlined in Table \ref{evaluation_criteria}.

The results, shown in Table \ref{table:scores}, indicate that FinRobot achieved strong scores for accuracy, with four experts awarding a perfect score of 10 and three others scoring it a 9. This suggests a high level of agreement among reviewers regarding the report's factual precision. Logical coherence also received high marks, although with slight variation. Two reviewers rated it a 10, four gave it a 9, and one awarded a 9.5, reflecting minor differences in perception of the report's structure and flow. The storytelling ability, while well-rated, showed potential areas for improvement, as some reviewers felt the narrative could be made even more engaging. Detailed comments and insights from the reviewers are provided in the supplementary materials.

\subsubsection{LLM Review} In addition to expert evaluations, we utilized GPT-4 to assess the report’s quality. GPT-4 was prompted with specific instructions to evaluate the report based on the same three dimensions: accuracy, logical coherence, and storytelling. The prompts used are illustrated in Fig. \ref{fig:prompt_template}. GPT-4’s evaluation aligned closely with the expert assessments, further validating the report’s strengths in factual accuracy, logical flow, and narrative engagement. Detailed output from GPT-4's assessment can be found in Fig. \ref{fig:gpt_ans1}.

\subsubsection{Stability Assessment} To ensure the robustness of FinRobot’s report generation, we conducted a stability assessment by generating multiple reports on the same topic and evaluating them for consistency. GPT-4 was tasked with scoring each report on accuracy, logicality, and storytelling, allowing us to measure consistency across different instances. Additionally, we generated contrast samples using zero-shot, few-shot, and chain-of-thought prompting, each evaluated on the same dimensions.

For each dimension, GPT-4 was given a targeted prompt, such as, “Evaluate the report based on logical coherence and assign a score from 0 to 10, with 10 representing flawless coherence.” This structured approach provided a systematic way to quantify report quality. Fig. \ref{fig:quality_analysis} presents a density plot comparing the scores of reports generated by FinRobot against those produced using zero-shot, few-shot, and chain-of-thought methods. As depicted, FinRobot consistently delivers high-quality reports that outperform other prompting methods in terms of accuracy, logical coherence, and storytelling, confirming the reliability and stability of its output.

\section{Conclusion}

This paper introduced FinRobot, an AI agent framework for high-quality, automated equity research. Leveraging a multi-agent Chain of Thought system, FinRobot integrates discretionary judgment with quantitative analysis, setting a new standard in AI-driven financial analysis. Its dynamic data pipeline and robust risk assessments ensure timely and balanced insights.

Future developments will expand FinRobot’s capabilities across various industries and asset classes, including comprehensive coverage of the Dow 30. Enhancements like reinforcement learning and sentiment analysis will further deepen FinRobot’s analytical power, offering financial professionals more versatile and innovative research tools.

\bibliographystyle{ACM-Reference-Format}
\bibliography{ref}


\begin{thebibliography}{23}


\ifx \showCODEN    \undefined \def \showCODEN     #1{\unskip}     \fi
\ifx \showDOI      \undefined \def \showDOI       #1{#1}\fi
\ifx \showISBNx    \undefined \def \showISBNx     #1{\unskip}     \fi
\ifx \showISBNxiii \undefined \def \showISBNxiii  #1{\unskip}     \fi
\ifx \showISSN     \undefined \def \showISSN      #1{\unskip}     \fi
\ifx \showLCCN     \undefined \def \showLCCN      #1{\unskip}     \fi
\ifx \shownote     \undefined \def \shownote      #1{#1}          \fi
\ifx \showarticletitle \undefined \def \showarticletitle #1{#1}   \fi
\ifx \showURL      \undefined \def \showURL       {\relax}        \fi
\providecommand\bibfield[2]{#2}
\providecommand\bibinfo[2]{#2}
\providecommand\natexlab[1]{#1}
\providecommand\showeprint[2][]{arXiv:#2}

\bibitem[Abarbanell and Bushee(1997)]%
        {abarbanell1997fundamental}
\bibfield{author}{\bibinfo{person}{Jeffrey~S Abarbanell} {and} \bibinfo{person}{Brian~J Bushee}.} \bibinfo{year}{1997}\natexlab{}.
\newblock \showarticletitle{Fundamental analysis, future earnings, and stock prices}.
\newblock \bibinfo{journal}{\emph{Journal of accounting research}} \bibinfo{volume}{35}, \bibinfo{number}{1} (\bibinfo{year}{1997}), \bibinfo{pages}{1--24}.
\newblock


\bibitem[Berman and Knight(2013)]%
        {berman2013financial}
\bibfield{author}{\bibinfo{person}{Karen Berman} {and} \bibinfo{person}{Joe Knight}.} \bibinfo{year}{2013}\natexlab{}.
\newblock \bibinfo{booktitle}{\emph{Financial intelligence, revised edition: A manager's guide to knowing what the numbers really mean}}.
\newblock \bibinfo{publisher}{Harvard Business Review Press}.
\newblock


\bibitem[Brown et~al\mbox{.}(2020)]%
        {brown2020language}
\bibfield{author}{\bibinfo{person}{Tom Brown}, \bibinfo{person}{Benjamin Mann}, \bibinfo{person}{Nick Ryder}, \bibinfo{person}{Melanie Subbiah}, \bibinfo{person}{Jared~D Kaplan}, \bibinfo{person}{Prafulla Dhariwal}, \bibinfo{person}{Arvind Neelakantan}, \bibinfo{person}{Pranav Shyam}, \bibinfo{person}{Girish Sastry}, \bibinfo{person}{Amanda Askell}, {et~al\mbox{.}}} \bibinfo{year}{2020}\natexlab{}.
\newblock \showarticletitle{Language models are few-shot learners}.
\newblock \bibinfo{journal}{\emph{Advances in Neural Information Processing Systems}}  \bibinfo{volume}{33} (\bibinfo{year}{2020}), \bibinfo{pages}{1877--1901}.
\newblock


\bibitem[Greenwald et~al\mbox{.}(2004)]%
        {greenwald2004value}
\bibfield{author}{\bibinfo{person}{B.C. Greenwald}, \bibinfo{person}{J. Kahn}, \bibinfo{person}{P.D. Sonkin}, {and} \bibinfo{person}{M. van Biema}.} \bibinfo{year}{2004}\natexlab{}.
\newblock \bibinfo{booktitle}{\emph{Value Investing: From Graham to Buffett and Beyond}}.
\newblock \bibinfo{publisher}{Wiley}.
\newblock
\showISBNx{9780471463399}
\showLCCN{2001017635}
\urldef\tempurl%
\url{https://books.google.com.sg/books?id=gvCzlskpZxoC}
\showURL{%
\tempurl}


\bibitem[Henrique et~al\mbox{.}(2019)]%
        {henrique2019literature}
\bibfield{author}{\bibinfo{person}{Bruno~Miranda Henrique}, \bibinfo{person}{Vinicius~Amorim Sobreiro}, {and} \bibinfo{person}{Herbert Kimura}.} \bibinfo{year}{2019}\natexlab{}.
\newblock \showarticletitle{Literature review: Machine learning techniques applied to financial market prediction}.
\newblock \bibinfo{journal}{\emph{Expert Systems with Applications}}  \bibinfo{volume}{124} (\bibinfo{year}{2019}), \bibinfo{pages}{226--251}.
\newblock


\bibitem[Huang et~al\mbox{.}(2023)]%
        {huang2023finbert}
\bibfield{author}{\bibinfo{person}{Allen~H Huang}, \bibinfo{person}{Hui Wang}, {and} \bibinfo{person}{Yi Yang}.} \bibinfo{year}{2023}\natexlab{}.
\newblock \showarticletitle{FinBERT: A large language model for extracting information from financial text}.
\newblock \bibinfo{journal}{\emph{Contemporary Accounting Research}} \bibinfo{volume}{40}, \bibinfo{number}{2} (\bibinfo{year}{2023}), \bibinfo{pages}{806--841}.
\newblock


\bibitem[Jiang(2021)]%
        {jiang2021applications}
\bibfield{author}{\bibinfo{person}{Weiwei Jiang}.} \bibinfo{year}{2021}\natexlab{}.
\newblock \showarticletitle{Applications of deep learning in stock market prediction: recent progress}.
\newblock \bibinfo{journal}{\emph{Expert Systems with Applications}}  \bibinfo{volume}{184} (\bibinfo{year}{2021}), \bibinfo{pages}{115537}.
\newblock


\bibitem[Kim et~al\mbox{.}(2024)]%
        {kim2024financial}
\bibfield{author}{\bibinfo{person}{Alex Kim}, \bibinfo{person}{Maximilian Muhn}, {and} \bibinfo{person}{Valeri~V Nikolaev}.} \bibinfo{year}{2024}\natexlab{}.
\newblock \showarticletitle{Financial Statement Analysis with Large Language Models}.
\newblock \bibinfo{journal}{\emph{Chicago Booth Research Paper Forthcoming, Fama-Miller Working Paper}} (\bibinfo{year}{2024}).
\newblock


\bibitem[Kumar et~al\mbox{.}(2022)]%
        {kumar2022systematic}
\bibfield{author}{\bibinfo{person}{Deepak Kumar}, \bibinfo{person}{Pradeepta~Kumar Sarangi}, {and} \bibinfo{person}{Rajit Verma}.} \bibinfo{year}{2022}\natexlab{}.
\newblock \showarticletitle{A systematic review of stock market prediction using machine learning and statistical techniques}.
\newblock \bibinfo{journal}{\emph{Materials Today: Proceedings}}  \bibinfo{volume}{49} (\bibinfo{year}{2022}), \bibinfo{pages}{3187--3191}.
\newblock


\bibitem[Medhat et~al\mbox{.}(2014)]%
        {medhat2014sentiment}
\bibfield{author}{\bibinfo{person}{Walaa Medhat}, \bibinfo{person}{Ahmed Hassan}, {and} \bibinfo{person}{Hoda Korashy}.} \bibinfo{year}{2014}\natexlab{}.
\newblock \showarticletitle{Sentiment analysis algorithms and applications: A survey}.
\newblock \bibinfo{journal}{\emph{Ain Shams engineering journal}} \bibinfo{volume}{5}, \bibinfo{number}{4} (\bibinfo{year}{2014}), \bibinfo{pages}{1093--1113}.
\newblock


\bibitem[Nabipour et~al\mbox{.}(2020)]%
        {nabipour2020deep}
\bibfield{author}{\bibinfo{person}{Mojtaba Nabipour}, \bibinfo{person}{Pooyan Nayyeri}, \bibinfo{person}{Hamed Jabani}, \bibinfo{person}{Amir Mosavi}, \bibinfo{person}{Ely Salwana}, {and} \bibinfo{person}{Shahab S}.} \bibinfo{year}{2020}\natexlab{}.
\newblock \showarticletitle{Deep learning for stock market prediction}.
\newblock \bibinfo{journal}{\emph{Entropy}} \bibinfo{volume}{22}, \bibinfo{number}{8} (\bibinfo{year}{2020}), \bibinfo{pages}{840}.
\newblock


\bibitem[Nie et~al\mbox{.}(2024)]%
        {nie2024survey}
\bibfield{author}{\bibinfo{person}{Yuqi Nie}, \bibinfo{person}{Yaxuan Kong}, \bibinfo{person}{Xiaowen Dong}, \bibinfo{person}{John~M Mulvey}, \bibinfo{person}{H~Vincent Poor}, \bibinfo{person}{Qingsong Wen}, {and} \bibinfo{person}{Stefan Zohren}.} \bibinfo{year}{2024}\natexlab{}.
\newblock \showarticletitle{A Survey of Large Language Models for Financial Applications: Progress, Prospects and Challenges}.
\newblock \bibinfo{journal}{\emph{arXiv preprint arXiv:2406.11903}} (\bibinfo{year}{2024}).
\newblock


\bibitem[Penman(2010)]%
        {penman2010accounting}
\bibfield{author}{\bibinfo{person}{S. Penman}.} \bibinfo{year}{2010}\natexlab{}.
\newblock \bibinfo{booktitle}{\emph{Accounting for Value}}.
\newblock \bibinfo{publisher}{Columbia University Press}.
\newblock
\showISBNx{9780231521857}
\showLCCN{2010042335}
\urldef\tempurl%
\url{https://books.google.com.sg/books?id=5A8fY-RKZLIC}
\showURL{%
\tempurl}


\bibitem[Radford et~al\mbox{.}(2018)]%
        {radford2018improving}
\bibfield{author}{\bibinfo{person}{Alec Radford}, \bibinfo{person}{Karthik Narasimhan}, \bibinfo{person}{Tim Salimans}, \bibinfo{person}{Ilya Sutskever}, {et~al\mbox{.}}} \bibinfo{year}{2018}\natexlab{}.
\newblock \showarticletitle{Improving language understanding by generative pre-training}.
\newblock \bibinfo{journal}{\emph{OpenAI}} (\bibinfo{year}{2018}).
\newblock


\bibitem[Sohangir et~al\mbox{.}(2018)]%
        {sohangir2018big}
\bibfield{author}{\bibinfo{person}{Sahar Sohangir}, \bibinfo{person}{Dingding Wang}, \bibinfo{person}{Anna Pomeranets}, {and} \bibinfo{person}{Taghi~M Khoshgoftaar}.} \bibinfo{year}{2018}\natexlab{}.
\newblock \showarticletitle{Big Data: Deep Learning for financial sentiment analysis}.
\newblock \bibinfo{journal}{\emph{Journal of Big Data}} \bibinfo{volume}{5}, \bibinfo{number}{1} (\bibinfo{year}{2018}), \bibinfo{pages}{1--25}.
\newblock


\bibitem[Subramanyam(2014)]%
        {subramanyam2014financial}
\bibfield{author}{\bibinfo{person}{KR Subramanyam}.} \bibinfo{year}{2014}\natexlab{}.
\newblock \bibinfo{booktitle}{\emph{Financial statement analysis}}.
\newblock \bibinfo{publisher}{McGraw-Hill}.
\newblock


\bibitem[Wei et~al\mbox{.}(2022)]%
        {wei2022chain}
\bibfield{author}{\bibinfo{person}{Jason Wei}, \bibinfo{person}{Xuezhi Wang}, \bibinfo{person}{Dale Schuurmans}, \bibinfo{person}{Maarten Bosma}, \bibinfo{person}{Fei Xia}, \bibinfo{person}{Ed Chi}, \bibinfo{person}{Quoc~V Le}, \bibinfo{person}{Denny Zhou}, {et~al\mbox{.}}} \bibinfo{year}{2022}\natexlab{}.
\newblock \showarticletitle{Chain-of-thought prompting elicits reasoning in large language models}.
\newblock \bibinfo{journal}{\emph{Advances in neural information processing systems}}  \bibinfo{volume}{35} (\bibinfo{year}{2022}), \bibinfo{pages}{24824--24837}.
\newblock


\bibitem[Wu et~al\mbox{.}(2023)]%
        {wu2023bloomberggpt}
\bibfield{author}{\bibinfo{person}{Shijie Wu}, \bibinfo{person}{Ozan Irsoy}, \bibinfo{person}{Steven Lu}, \bibinfo{person}{Vadim Dabravolski}, \bibinfo{person}{Mark Dredze}, \bibinfo{person}{Sebastian Gehrmann}, \bibinfo{person}{Prabhanjan Kambadur}, \bibinfo{person}{David Rosenberg}, {and} \bibinfo{person}{Gideon Mann}.} \bibinfo{year}{2023}\natexlab{}.
\newblock \showarticletitle{{BloombergGPT}: A large language model for finance}.
\newblock \bibinfo{journal}{\emph{arXiv preprint arXiv:2303.17564}} (\bibinfo{year}{2023}).
\newblock


\bibitem[Yang et~al\mbox{.}(2023)]%
        {yang2023fingpt}
\bibfield{author}{\bibinfo{person}{Hongyang Yang}, \bibinfo{person}{Xiao-Yang Liu}, {and} \bibinfo{person}{Christina~Dan Wang}.} \bibinfo{year}{2023}\natexlab{}.
\newblock \showarticletitle{FinGPT: Open-Source Financial Large Language Models}.
\newblock \bibinfo{journal}{\emph{FinLLM Symposium at IJCAI 2023}} (\bibinfo{year}{2023}).
\newblock


\bibitem[Yu et~al\mbox{.}(2023)]%
        {yu2023finmem}
\bibfield{author}{\bibinfo{person}{Yangyang Yu}, \bibinfo{person}{Haohang Li}, \bibinfo{person}{Zhi Chen}, \bibinfo{person}{Yuechen Jiang}, \bibinfo{person}{Yang Li}, \bibinfo{person}{Denghui Zhang}, \bibinfo{person}{Rong Liu}, \bibinfo{person}{Jordan~W. Suchow}, {and} \bibinfo{person}{Khaldoun Khashanah}.} \bibinfo{year}{2023}\natexlab{}.
\newblock \bibinfo{title}{FinMem: A Performance-Enhanced LLM Trading Agent with Layered Memory and Character Design}.
\newblock
\newblock
\showeprint[arxiv]{2311.13743}~[q-fin.CP]


\bibitem[Zhang et~al\mbox{.}(2023)]%
        {zhang2023fingptrag}
\bibfield{author}{\bibinfo{person}{Boyu Zhang}, \bibinfo{person}{Hongyang Yang}, \bibinfo{person}{Tianyu Zhou}, \bibinfo{person}{Ali Babar}, {and} \bibinfo{person}{Xiao-Yang Liu}.} \bibinfo{year}{2023}\natexlab{}.
\newblock \showarticletitle{Enhancing Financial Sentiment Analysis via Retrieval Augmented Large Language Models}.
\newblock \bibinfo{journal}{\emph{ACM International Conference on AI in Finance (ICAIF)}} (\bibinfo{year}{2023}).
\newblock


\bibitem[Zhang et~al\mbox{.}(2024)]%
        {zhang2024finagent}
\bibfield{author}{\bibinfo{person}{Wentao Zhang}, \bibinfo{person}{Lingxuan Zhao}, \bibinfo{person}{Haochong Xia}, \bibinfo{person}{Shuo Sun}, \bibinfo{person}{Jiaze Sun}, \bibinfo{person}{Molei Qin}, \bibinfo{person}{Xinyi Li}, \bibinfo{person}{Yuqing Zhao}, \bibinfo{person}{Yilei Zhao}, \bibinfo{person}{Xinyu Cai}, {et~al\mbox{.}}} \bibinfo{year}{2024}\natexlab{}.
\newblock \showarticletitle{FinAgent: A Multimodal Foundation Agent for Financial Trading: Tool-Augmented, Diversified, and Generalist}.
\newblock \bibinfo{journal}{\emph{arXiv preprint arXiv:2402.18485}} (\bibinfo{year}{2024}).
\newblock


\bibitem[Zhao et~al\mbox{.}(2023)]%
        {zhao2023survey}
\bibfield{author}{\bibinfo{person}{Wayne~Xin Zhao}, \bibinfo{person}{Kun Zhou}, \bibinfo{person}{Junyi Li}, \bibinfo{person}{Tianyi Tang}, \bibinfo{person}{Xiaolei Wang}, \bibinfo{person}{Yupeng Hou}, \bibinfo{person}{Yingqian Min}, \bibinfo{person}{Beichen Zhang}, \bibinfo{person}{Junjie Zhang}, \bibinfo{person}{Zican Dong}, {et~al\mbox{.}}} \bibinfo{year}{2023}\natexlab{}.
\newblock \showarticletitle{A survey of large language models}.
\newblock \bibinfo{journal}{\emph{arXiv preprint arXiv:2303.18223}} (\bibinfo{year}{2023}).
\newblock


\end{thebibliography}

\newpage

\appendix

\begin{figure*}
\centering
\includegraphics[scale = 0.8]{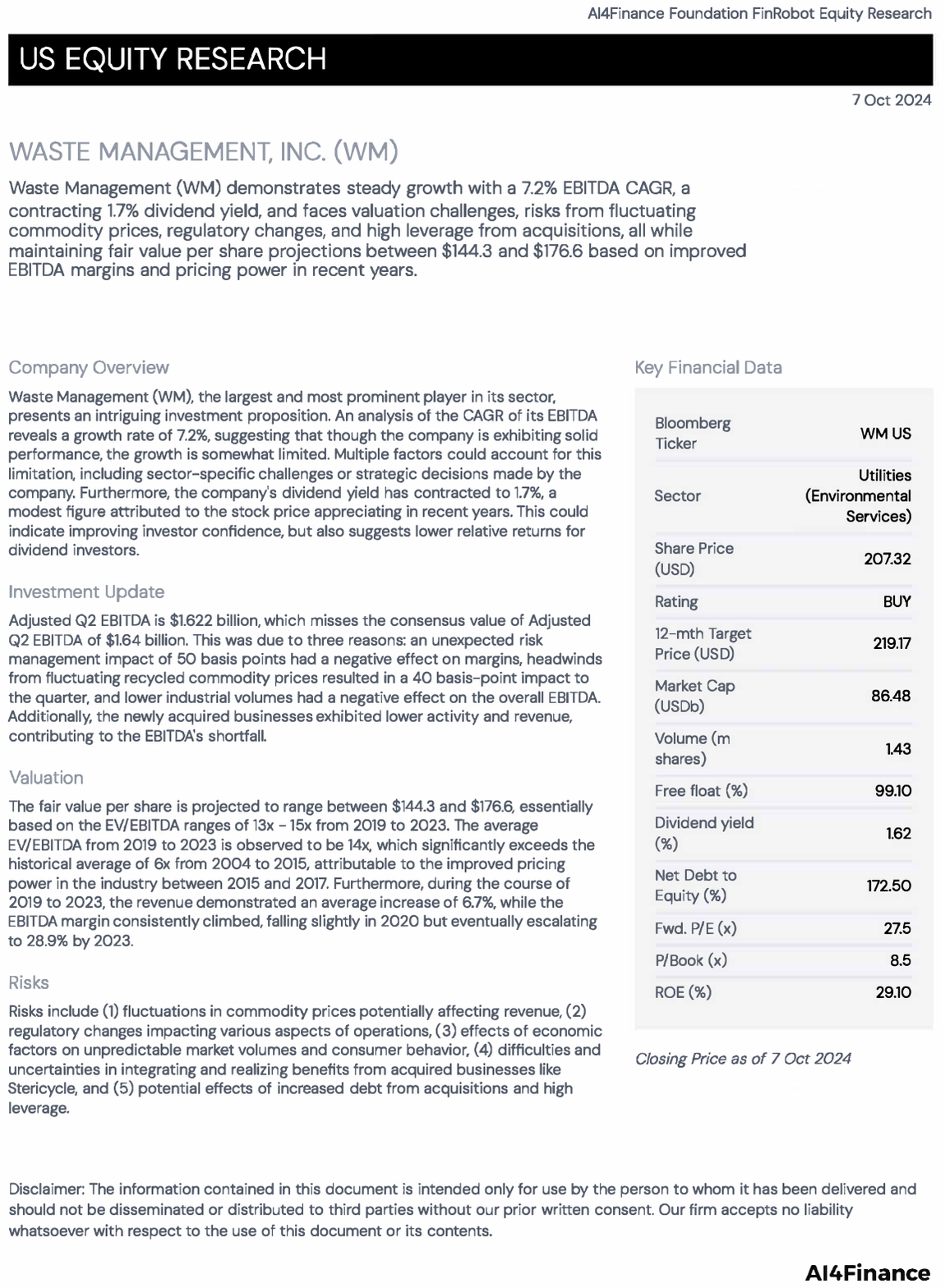}
\label{fig:wm_report1}
\vspace{-1mm}
\end{figure*}

\begin{figure*}
\centering
\includegraphics[scale = 0.88]{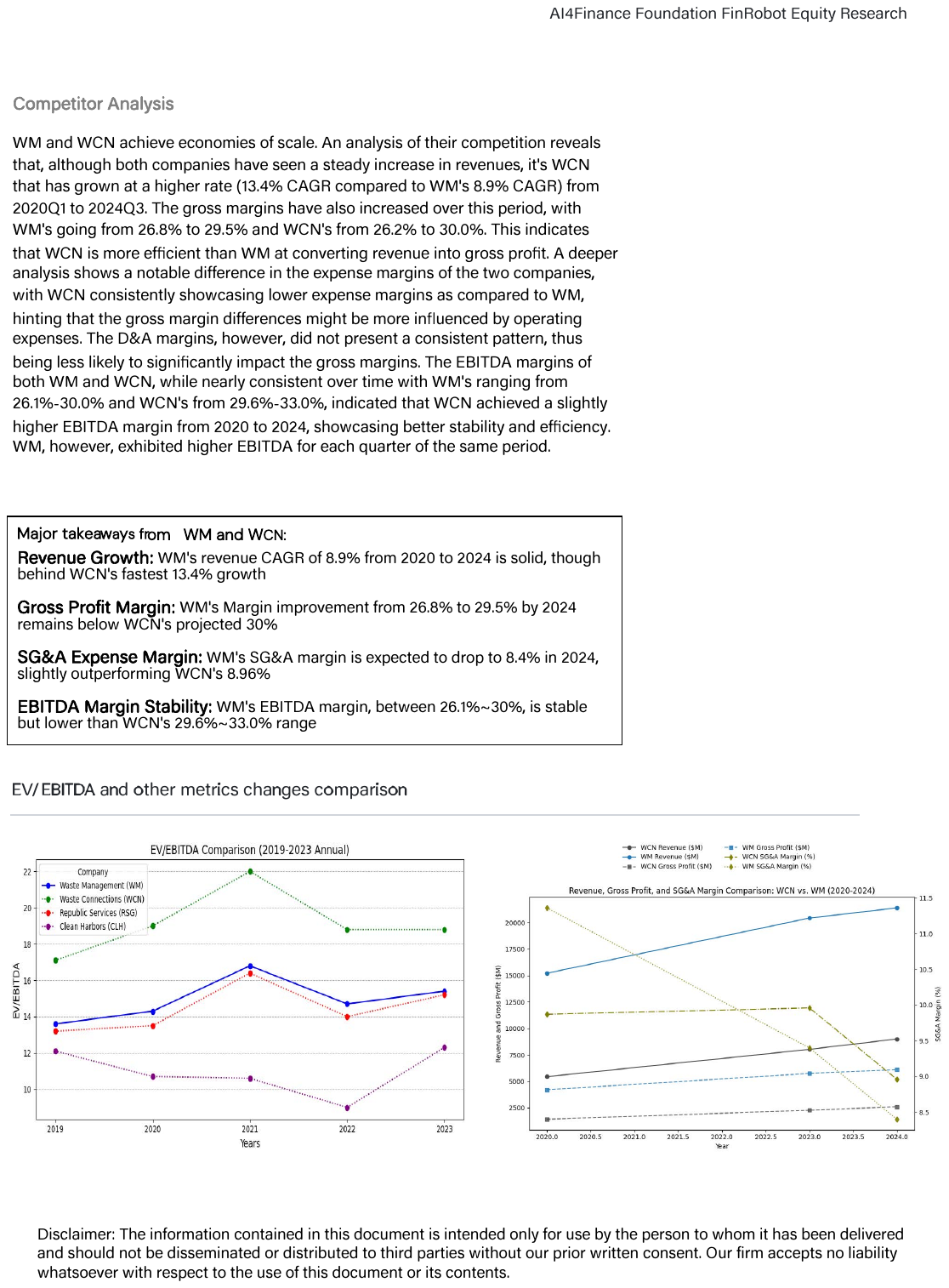}
\label{fig:wm_report1}
\vspace{-1mm}
\end{figure*}

\begin{figure*}
\centering
\includegraphics[scale = 0.88]{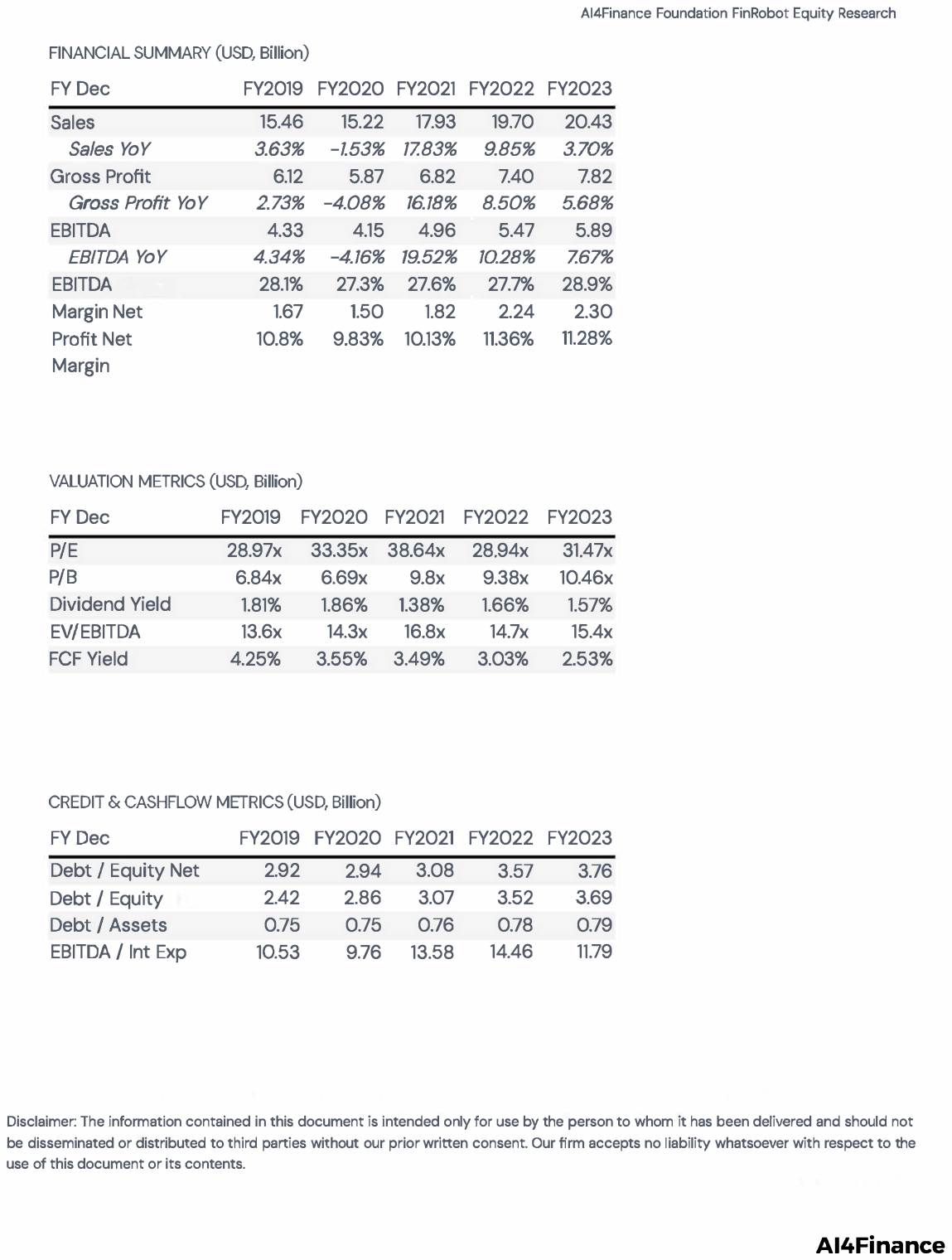}
\label{fig:wm_report1}
\vspace{-1mm}
\end{figure*}

\appendix
\clearpage
\section{Supplementary Data}

\begin{center}
\begin{minipage}[t]{0.4\textwidth}
\end{minipage}%
\hfill
\begin{minipage}[t]{0.55\textwidth}
\begin{table}[H] 
\centering
\begin{tabular}{|c|p{1.5\textwidth}|}
\hline
\textbf{Reviewer} & \textbf{Reviewer Score Comments} \\
\hline
Reviewer 1 & The generated report appears very accurate, with exact numbers for past financials, current projections, and valuation metrics. The report contains all the metrics and numbers critical for analyzing an equity research report. The report generally flows logically, though some of the wording is slightly unnatural. There is a strong focus on quantitative data, but the impact or importance of that data is not always clear. Sometimes it reads more like a list of statistics than an analytical report. Despite this, it is visually appealing and well-organized. \\
\hline
Reviewer 2 & This is an excellent overview of Waste Management that covers all of the basics that I could want to know as an equity investor. The information is presented clearly and is well organized. The writing skills displayed by this LLM make the report exceedingly easy to follow. I could definitely see myself using this tool the next time I want to get some preliminary information about a company.\\
\hline
Reviewer 3 & For someone reviewing the information for the first time, the data appears reasonable. Nothing overly conspicuous or eyebrow-raising. The overall layout of the document is very coherent and the headings make logical sense. The storytelling is solid, with some room for modest improvements. Under the "Company Overview" section, there should be a more clear description of what Waste Management does and what its sector is; even if it may seem self-evident given the name of the company, you should clarify this information since there are many cases in which the answer is less obvious. Furthermore, the sentence structure of the document is a bit awkward and repetitive. I'd lay off on the appositives and subordinate clauses to make the text cleaner and more direct. Finally, the thesis could be clearer; your opening paragraph could do a better job at suggesting that WM is a BUY. But overall, not bad for AI -- it's only going to get better with the exponential improvements in LLMs that we're witnessing!\\
\hline
Reviewer 4 & This model is capable of compiling a great deal of useful information together in a concise manner. While highly useful for run-of-the-mill equities or getting a quick report, qualitative analysis of blue chip news driven stocks is still best left to humans.\\
\hline
Reviewer 5 & The research report flows smoothly for the first read-through, the valuation and proposition make sense logically and coherently. Upon second-glance there were minor errors but, minimal in the scope of the document which presented ideas clearly and concisely. As story telling goes, it is difficult to judge knowing it is an AI stock pitch, if I didn't know maybe I would grade differently.\\
\hline
Reviewer 6 & The paper is very well written covering all aspects of WM. Data and statistics have been used very effectively to tell the story about WM and the future of the stock. The paper is well rounded with the potential risks considered after talking about the future valuation of the stock. There might be very minimal inaccuracy in valuation due to lack of multiple methods but overall extremely well written paper.\\
\hline
Reviewer 7 & All numbers are accurate and calculations are correct. All relevant metrics are present and easy to comprehend and digest. Grammar is correct and sentences are logical. Easy to get a picture of WM financials and begin to analyze the company from the data and analysis provided. Structure of the report is well thought out and industry standard. Reading the headline of each subsection provides the reader with general indication of the material discussed. Key Financial Data is easy to find and synthesize thanks to different color background. Logic could be slightly improved by shortening isolated sentences that border on run-ons by creating multiple sentences for concision. Introductory section touches on the main attributes of the company, including discussion of both positive attributes and risks. General feelings about the company can be felt just by reading the introduction. Structure of subsections makes sense, and concision promotes digestibility of material for all readers. Voice seems occasionally repetitive and monotone, and the report could feel more engaging if the voice was more varied and engaging. Story could also be enhanced slightly if a short investment thesis is included somewhere in the analysis.
\\
\hline
\end{tabular}
\caption{Reviewer Score Comments for Equity Research Report}
\label{table:comments}
\end{table}
\end{minipage}
\end{center}

\end{document}